\documentclass[journal]{IEEEtran}

\usepackage{cite}
\usepackage{algorithmic}
\usepackage{array}
\usepackage{bbm}
\usepackage{amsfonts}
\usepackage{graphicx}
\usepackage[hidelinks]{hyperref}
\usepackage{xcolor}
\hypersetup{
	colorlinks  = true,
	linkcolor	={red!100!black},
	citecolor	={blue!100!black},
	urlcolor	={blue!100!black}
}
\usepackage{mathtools}
\usepackage{subfigure,epsfig}
\usepackage{amsmath, amsthm, amssymb}
\usepackage[outdir=./]{epstopdf}
\usepackage{dsfont}
\usepackage{mathrsfs}

\newtheorem{Corollary}{Corollary}

\def\figref#1{Fig.\,\ref{#1}}%
\newlength{\figwidth}
\begin{document}
\title{Handover Rate and Sojourn Time Analysis in Mobile Drone-Assisted Cellular Networks}
\author{Mohammad Salehi and Ekram Hossain
\thanks{The authors are with the 
Department of Electrical and Computer Engineering, 
University of Manitoba, Canada (Email: salehim@myumanitoba.ca, Ekram.Hossain@umanitoba.ca).
This work was supported by a Discovery Grant from the Natural Sciences and Engineering Research Council of Canada (NSERC).
} }

\maketitle

\begin{abstract}
To improve capacity and overcome some of the limitations of cellular wireless networks, drones with aerial base stations can be deployed to assist the terrestrial cellular wireless networks. The mobility of drones allows flexible network reconfiguration to adapt to dynamic traffic and channel conditions. However, this is achieved at the expense of more handovers since even a static user may experience a handover when the drones are mobile. In this letter, we provide an exact analysis of the handover rate and sojourn time (time between two subsequent handovers) for a network of drone base stations. We also show that among different speed distributions with the same mean, the handover rate is minimum when all drone base stations move with same speed.
\end{abstract}

\section{Introduction}
Using drones as aerial base stations, to assist terrestrial communications, is a promising approach to tackle the 5G and beyond 5G challenges. Owing to their mobility and high flexibility, drones can provide on-demand communications to ground users \cite{Mozaffari2019} but the downside is that 5G and beyond wireless networks with drone-assisted communications suffer from increased handover rate. In this letter, we derive the handover rate (mean number of handovers in a unit time) and mean sojourn time (mean service time by each drone base station) of a network of drone base stations.

Existing works study the impact of mobility on handover mostly through handover rate, handover probability, and sojourn time. In this regard, \cite{Lin2013} has derived the handover rate and mean sojourn time for a single-tier cellular network with Poisson point process (PPP) distributed terrestrial base stations. For multi-tier networks, \cite{Bao2015} has studied the handover rate, and \cite{salehi2020stochastic} has studies the mean sojourn time. Handover probability is also studied in \cite{Sadr} for single-tier networks and in \cite{Hsueh} for multi-tier networks. In terrestrial networks, handover occurs as the user moves across a cell boundary; however, in drone-assisted communications, even a static user may experience a handover due to drones' mobility. Therefore, it is critical to study the handover rate for these networks.

To study the handover rate in the context of drone-assisted cellular communication, most of the existing works only consider the scenario where the drones act as users, i.e., they focus on the connection between a drone and terrestrial base stations \cite{Amer2020,Arshad2019}. In this scenario, handover analysis is similar to the previous works except that instead of moving in two dimensions, drones can move in three dimensions. In this case, we need to project drones mobility onto $\mathbb{R}^2$ plane. Recently, \cite{Banagar2020} has studied the handover probability for a network of drone base stations\footnote{Definition of handover probability in \cite{Sadr,Hsueh} is different from \cite{Banagar2020}. To analyze the handover probability \cite{Sadr,Hsueh} only consider two time instants; for example, time 0 and $t$. On the other hand, \cite{Banagar2020} considers the entire time interval between 0 and $t$, i.e., $[0,t]$. To understand the difference and relation between these two, refer to \cite{Tabassum,salehi2020stochastic}.}. They have derived the exact result for the scenario where all drones move with the same speed, but, for the scenario with different speeds, only a bound is provided. However, {\em as the first step toward handover rate analysis, we need to derive the exact results. For this, we convert drones' mobility to user's mobility, so that instead of having multiple mobile nodes we only have a single mobile node.} Then, from the exact results, we can derive the handover rate and mean sojourn time following the same steps as in \cite{salehi2020stochastic}.

In Section II, we introduce the system model and state the methodology of analysis. Section III provides the analytical results. Numerical results are validated in Section IV, where we also study the effect of speed distribution for the drones. Finally, Section V concludes the paper. 

\section{System Model and Methodology of Analysis}
Consider a network of mobile drone base stations (BSs) that serves ground users. Drone BSs are initially (at time 0) distributed at height $h$ according to a two-dimensional homogeneous Poisson point process (PPP) $\Phi$ of density $\lambda$. Each drone BS moves in $z=h$ plane in a random direction $\theta$ with respect to the positive $x$-axis with velocity $v$ independent from other drone BSs and its location. $f_{\Theta}(.)$ and $f_V(.)$ provide distributions of $\theta$ and $v$. Assuming straight line trajectories for the drones provides performance bound for more complicated models. It also complies with drones' mobility model in 3GPP simulations \cite{3GPP}. From the stationarity of the homogeneous PPP, we can assume that a user is located at the origin of our coordination system. This user is always associated with the drone BS that provides the maximum averaged received power (i.e., nearest drone). Since all drone BSs are at the same height, for the purpose of handover rate and sojourn time analysis, we can ignore $h$ and focus on the $\mathbb{R}^2$ plane. Therefore, in the following section, by location we mean the projected location of the drone BS onto the $x$-$y$ plane.

The methodology of analysis of handover rate and sojourn time is as follows:
\begin{itemize}
    \item Step 1: We derive the conditional distribution of the sojourn time for the initially serving drone base station, given that this drone moves with velocity $v_0$ in direction $\theta_0$ (Section III.A). 
    \item Step 2: We calculate the handover rate from a drone base station with velocity $v_0$ and movement direction $\theta_0$ to any other drone base station (Section III.B).
    \item Step 3: Finally, the handover rate (inverse of mean sojourn time) can be obtained by integrating over different $v_0$, and $\theta_0$ (Section III.B). 
\end{itemize}

\section{Handover Rate and Sojourn Time Analysis}
The initial serving drone BS moves with velocity $v_0$ in direction $\theta_0$ with probability 
\begin{IEEEeqnarray}{rCl}
    \mathbb{P}(v_0,\theta_0)=f_V(v_0) {\rm d}v_0 f_{\Theta}(\theta_0){\rm d}\theta_0.
    \label{eq:PDF_tier}
\end{IEEEeqnarray}
Let us denote the distance between the (projected) location of the initial serving drone BS and the origin at time $0$ by $r_0$. Conditional probability density function (PDF) of $r_0$ given $v_0$ and $\theta_0$ is
\begin{IEEEeqnarray}{rCl}
    f_R(r_0 \mid v_0, \theta_0)=2\lambda\pi r_0 e^{-\lambda \pi r_0^2}.
    \label{eq:Conditional_PDF_serving_link_distance}
\end{IEEEeqnarray}
Since the model is isotropic (invariant under rotation), we can assume that (projected) location of the serving drone BS at time 0 is at $[r_0,0]^T$. In this section, we first derive the distribution of the time until the first handover, denoted by $\tilde{S}$, given that the initial serving drone BS moves with velocity $v_0$ in direction $\theta_0$. Then following the same steps as in \cite{salehi2020stochastic}, we derive the handover rate and sojourn time.

\subsection{Conditional Distribution of $\tilde{S}$}
The complementary cumulative distribution function (CCDF) of the time until the first handover $\tilde{S}$ given $r_0$, $v_0$, and $\theta_0$ can be obtained by
\allowdisplaybreaks{
\begin{IEEEeqnarray}{rCl}
    \IEEEeqnarraymulticol{3}{l}{ 
    \mathbb{P}\left( \tilde{S}>s \mid r_0, v_0, \theta_0 \right)= 
    \mathbb{P}\bigg( \bigcap\limits_{v} \bigcap\limits_{\theta}  \Phi_{v,\theta}^{(t)}\left(b\left([0,0]^T, r_0(t)\right)\right)=0   }
    \nonumber \\
    && \qquad \qquad \qquad \qquad \qquad \qquad \qquad  
    \forall t \in(0,s] \mid r_0, v_0, \theta_0 \bigg), 
    \label{eq:step1}
\end{IEEEeqnarray}}
where $r_0(t)=\sqrt{r_0^2+v_0^2t^2+2r_0v_0t\cos(\theta_0)}$ is the distance between the projected location of the initial serving drone BS and the origin at time $t$. $b\left([0,0]^T, r\right)$ denotes a ball with radius $r$ centered at the origin. $\Phi_{v,\theta}^{(0)} \subset \Phi \setminus \{[r_0,0]^T\}$ denotes the initial location of the non-serving drone BSs that move with velocity $v$ in direction $\theta$. According to the thinning property, their spatial distribution follows a PPP with density $\lambda f_V(v){\rm d}v f_{\Theta}(\theta){\rm d}\theta$ in $\mathbb{R}^2 \setminus b\left([0,0]^T,r_0\right)$. At time $t$, we have $\Phi_{v,\theta}^{(t)}=\Phi_{v,\theta}^{(0)}+vt[\cos(\theta), \sin(\theta)]^T$. Since $\Phi_{v,\theta}^{(0)}$s are independent for different $v$ and $\theta$, \eqref{eq:step1} can be further simplified as  
\begin{IEEEeqnarray}{rCl}
    \IEEEeqnarraymulticol{3}{l}{
    \mathbb{P}\left( \tilde{S}>s \mid r_0, v_0, \theta_0 \right)= \prod\limits_{v} \prod\limits_{\theta} \mathbb{P}\bigg( \Phi_{v,\theta}^{(0)}\left(b\left(\text{x}_{v,\theta}(t), r_0(t)\right)\right)=0} \nonumber \\
    && \qquad \qquad \qquad \qquad \qquad \qquad \qquad 
    \forall t \in(0,s] \mid r_0, v_0, \theta_0 \bigg), 
    \label{eq:step2}
\end{IEEEeqnarray}
where $\text{x}_{v,\theta}(t)=vt[\cos(\pi+\theta), \sin(\pi+\theta)]^T$. From \eqref{eq:step2} we can understand that sojourn time analysis for a static user in a network of moving drone BSs with velocity $v$ and direction $\theta$ is similar to the sojourn time analysis for a mobile user with velocity $v$ and direction $\pi+\theta$ in a network of static drone BSs. To calculate the right hand side of \eqref{eq:step2}, let us define 
\begin{IEEEeqnarray}{rCl}
    \mathcal{A}(v,\theta,r_0, v_0, \theta_0, s) \triangleq \left\{ \bigcup_{t} b\left(\text{x}_{v,\theta}(t), r_0(t)\right) : t\in[0,s]  \right\}. \IEEEeqnarraynumspace
    \label{eq:def_A}
\end{IEEEeqnarray}
Using the above definition, we can write
\begin{IEEEeqnarray}{rCl}
    \IEEEeqnarraymulticol{3}{l}{
    \mathbb{P}\left( \tilde{S}>s \mid r_0, v_0, \theta_0 \right)=} \nonumber \\
    \prod\limits_{v} \prod\limits_{\theta} \mathbb{P}\Big( \Phi_{v,\theta}^{(0)}\left( \mathcal{A}(v,\theta,r_0, v_0, \theta_0,s)\setminus b\left([0,0]^T,r_0\right) \right)=0 && \nonumber  \\
     \qquad \qquad \qquad \qquad \mid r_0, v_0, \theta_0 \Big). && \IEEEeqnarraynumspace
    \label{eq:step3}
\end{IEEEeqnarray}
The above probability can be calculated by using the void probability of the PPP. 
\begin{IEEEeqnarray}{rCl}
    \IEEEeqnarraymulticol{3}{l}{
    \mathbb{P}\left( \tilde{S}>s \mid r_0, v_0, \theta_0 \right)} \nonumber \\
    &\overset{\text{(a)}}{=}& \prod\limits_{v} \prod\limits_{\theta} e^{-\lambda \left(|\mathcal{A}(v,\theta,r_0, v_0, \theta_0,s)|-\pi r_0^2\right)f_V(v){\rm d}v f_{\Theta}(\theta){\rm d}\theta} \nonumber \\
    &=& e^{-\lambda \int\limits_v \int\limits_{\theta} |\mathcal{A}(v,\theta,r_0, v_0, \theta_0,s)| f_V(v) f_{\Theta}(\theta){\rm d}\theta{\rm d}v+\lambda\pi r_0^2}. 
    \label{eq:step4}
\end{IEEEeqnarray}
where $|\mathcal{A}|$ denotes area of the region $\mathcal{A}$, and (a) follows from $b\left([0,0]^T,r_0\right) \subset \mathcal{A}(v,\theta,r_0, v_0, \theta_0,s)$. Note that $|\mathcal{A}(v,\theta,r_0, v_0, \theta_0,s)|$ does not depend on $\theta$; thus, we can write
\begin{IEEEeqnarray}{rCl}
    \mathbb{P}\left( \tilde{S}>s \mid r_0, v_0, \theta_0 \right)
    &=& e^{-\lambda \int\limits_v |\mathcal{A}(v,r_0, v_0, \theta_0,s)| f_V(v) {\rm d}v+\lambda\pi r_0^2}, \IEEEeqnarraynumspace
    \label{eq:step5}
\end{IEEEeqnarray}
where 
\begin{IEEEeqnarray}{rCl}
    \mathcal{A}(v,r_0, v_0, \theta_0,s) = \left\{ \bigcup_{t} b\left([vt,0]^T, r_0(t)\right) : t\in[0,s]  \right\}. \IEEEeqnarraynumspace
    \label{eq:def_A_modified}
\end{IEEEeqnarray}
We can calculate $|\mathcal{A}(v,r_0, v_0, \theta_0,s)|$ from \textbf{Theorem 1} in \cite{salehi2020stochastic} by changing $r_0 \to \frac{v}{v_0}r_0$, $\beta_{kj} \to \frac{v}{v_0}$, $\theta \to \pi+\theta_0$, and $T \to s$. Using \eqref{eq:Conditional_PDF_serving_link_distance} yields 
\begin{IEEEeqnarray}{rCl}
    \IEEEeqnarraymulticol{3}{l}{
    \mathbb{P}\left( \tilde{S}>s \mid v_0, \theta_0 \right)=} \nonumber \\
    && \int_0^{\infty} 2 \lambda \pi r_0
    e^{-\lambda \int\limits_v |\mathcal{A}(v,r_0, v_0, \theta_0,s)| f_V(v) {\rm d}v} {\rm d}r_0. 
    \label{eq:main_result_subsection1}
\end{IEEEeqnarray}

{\em Remark}: To derive the distribution of $\tilde{S}$ for a static user in a hybrid network with terrestrial BSs and mobile drones\footnote{For this hybrid network, we only derive the distribution of $\tilde{S}$ since this is the fundamental step. Other steps are straightforward as will be discussed in the next subsection.}, let us use subscript 1 for the tier of drone BSs and subscript 2 for the tier of terrestrial BSs. $B_i$, $h_i$, $\lambda_i$, and $\alpha_i$ denote the bias factor\footnote{Bias factor also incorporates the effect of transmission power and mean power of small scale fading.}, height, density, and path-loss exponent of tier $i$, where $i\in\{1,2\}$. Let us define 
\begin{IEEEeqnarray}{rCl}
    f_{i,j}(x)=\sqrt{ 
    \left[ \left(\frac{B_i}{B_j}\right)^{2/\alpha_i} \left(r_j^2+h_j^2\right)^{\alpha_j/\alpha_i}-h_i^2 \right]^+}, \nonumber
\end{IEEEeqnarray}
where $i,j\in\{1,2\}$ and $[y]^+=\max(0,y)$. With maximum biased averaged receive power association, the user is initially served by a tier $j$ BS if $r_i>f_{i,j}(r_j)$, $i \neq j$, where $r_i$ is the projected distance of the nearest tier $i$ BS to the origin at time 0. Due to independence of drone BSs and terrestrial BSs, when the user is initially served by a drone BS, we have
\begin{IEEEeqnarray}{rCl}
    \IEEEeqnarraymulticol{3}{l}{ 
    \mathbb{P}\left( \tilde{S}>s \mid r_0, v_0, \theta_0, \text{tier}=1 \right)} \nonumber \\
    &=& e^{-\lambda_1 \int\limits_v |\mathcal{A}(v,r_0, v_0, \theta_0,s)| f_V(v) {\rm d}v+\lambda_1\pi r_0^2} \nonumber \\
    &&\times\, e^{-\lambda_2 \pi \left[f_{2,1}^2(\max(r_0,r_0(s))-f_{2,1}^2(r_0)\right]}. \nonumber
\end{IEEEeqnarray}
When the serving BS at time 0 is a terrestrial BS, we have
\begin{IEEEeqnarray}{rCl}
    \mathbb{P}\left( \tilde{S}>s \mid r_0, \text{tier}=2 \right) = 
    e^{-2\lambda_1 \mathbb{E}[v] s f_{1,2}(r_0)}. \nonumber
\end{IEEEeqnarray}

\subsection{Main Results}
To derive the handover rate and mean sojourn time, we first need to calculate \cite{salehi2020stochastic}
\begin{IEEEeqnarray}{rCl}
    \mathbb{E}[L \mid v_0, \theta_0] = \lim_{z \to 0} \frac{z}{1-\mathbb{P}\left( \tilde{S}>\frac{z}{v_0} \mid v_0, \theta_0 \right)}.
\end{IEEEeqnarray}
For a drone BS with velocity $v_0$ and movement direction $\theta_0$, $\mathbb{E}[L \mid v_0, \theta_0]$ is the average length of its trajectory during which the drone BS serves the user at the origin. Thus, mean sojourn time for drone BSs that move with velocity $v_0$ in direction $\theta_0$ is 
\begin{IEEEeqnarray}{rCl}
    \mathbb{E}[S \mid v_0, \theta_0] &=& \frac{\mathbb{E}[L \mid v_0, \theta_0]}{v_0} \nonumber \\
    &\stackrel{\text{(a)}}{=}& \frac{1}{v_0} \times
        \frac{-1}
        {\frac{{\rm d}}{{\rm d}z}\mathbb{P}\left( \tilde{S}>\frac{z}{v_0} \mid v_0, \theta_0 \right) \Big|_{z=0}}, 
    \label{eq:mean-sojounr-time-definition}
\end{IEEEeqnarray}
where (a) follows from the L'Hospital's Rule. From \eqref{eq:main_result_subsection1}, we have
\allowdisplaybreaks{
\begin{IEEEeqnarray}{rCl}
    \IEEEeqnarraymulticol{3}{l}{\frac{{\rm d}}{{\rm d}z}\mathbb{P}\left( \tilde{S}>\frac{z}{v_0} \mid v_0, \theta_0 \right) =  } 
    \nonumber \\
    && - \int_0^{\infty} 2 \lambda \pi r_0 e^{-\lambda \int\limits_v |\mathcal{A}(v,r_0, v_0, \theta_0,\frac{z}{v_0})| f_V(v) {\rm d}v} 
    \nonumber \\
    &&\quad \times \left( \lambda \int\limits_v \frac{{\rm d}}{{\rm d}z} |\mathcal{A}(v,r_0, v_0, \theta_0,\frac{z}{v_0})| f_V(v) {\rm d}v \right) {\rm d}r_0.
    \label{eq:denominator}
\end{IEEEeqnarray}}
Denominator of \eqref{eq:mean-sojounr-time-definition} can be calculated by substituting $|\mathcal{A}(v,r_0, v_0, \theta_0, 0)|=\pi r_0^2$, and 
\begin{IEEEeqnarray}{rCl}
    \IEEEeqnarraymulticol{3}{l}{
    \frac{{\rm d}}{{\rm d}z} |\mathcal{A}(v,r_0, v_0, \theta_0,\frac{z}{v_0})| \Big|_{z=0} = 2r_0 \times}
    \nonumber \\
    \begin{cases}
        \sqrt{ \left(\frac{v}{v_0}\right)^2-\cos^2\theta_0 } + \cos\theta_0 \cos^{-1}\left(-\frac{\cos\theta_0}{\frac{v}{v_0}} \right),
        \\
        \qquad \qquad \qquad \qquad \qquad \qquad \qquad \qquad
        \text{if   } |\cos\theta_0| \le \frac{v}{v_0}, 
        \\
        0, \qquad \qquad \qquad \qquad \qquad \qquad \qquad \quad \: \!
        \text{if  } \frac{v}{v_0} < -\cos\theta_0,
        \\
        \pi \cos\theta_0, \qquad \qquad \qquad \qquad \qquad \qquad \, \: 
        \text{if  } \frac{v}{v_0} < \cos\theta_0,
    \end{cases}. \nonumber
\end{IEEEeqnarray}
in \eqref{eq:denominator}. Therefore, 
\begin{IEEEeqnarray}{rCl}
    \mathbb{E}[S \mid v_0, \theta_0] = \frac{1}{\sqrt{\lambda}  \mathbb{E}_v[\mathcal{F}(v,v_0,\theta_0)]},
    \label{eq:ES_for_v0_thta0}
\end{IEEEeqnarray}
where $\mathbb{E}_v$ denotes the expectation with respect to $v$, and
\begin{IEEEeqnarray}{rCl}
    \IEEEeqnarraymulticol{3}{l}{
    \mathcal{F}(v,v_0,\theta_0) =  \pi v_0\cos\theta_0 \mathbf{1}\left(  \frac{v}{v_0} < \cos\theta_0 \right) } 
    \nonumber \\
    && + v_0 \Bigg( \sqrt{ \left(\frac{v}{v_0}\right)^2-\cos^2\theta_0 } 
    + \cos\theta_0 \cos^{-1}\left(-\frac{\cos\theta_0}{\frac{v}{v_0}} \right)\Bigg) 
    \nonumber \\
    &&\times\, \mathbf{1}\left( |\cos\theta_0| \le \frac{v}{v_0} \right). 
    \label{eq:F}
\end{IEEEeqnarray}
$\mathbf{1}(.)$, in \eqref{eq:F}, is the indicator function. From \eqref{eq:ES_for_v0_thta0}, we can calculate the mean number of handovers from (to) a drone BS with velocity $v_0$ and movement direction $\theta_0$ to (from) any other drone BS as \cite{salehi2020stochastic}
\begin{IEEEeqnarray}{rCl}
    H_{v_0,\theta_0} &=& \frac{\mathbb{P}(v_0,\theta_0)}
                            {\mathbb{E}[S \mid v_0, \theta_0]} \nonumber \\
                     &=& \sqrt{\lambda}  \mathbb{E}_v[\mathcal{F}(v,v_0,\theta_0)]
                     f_V(v_0) {\rm d}v_0 f_{\Theta}(\theta_0){\rm d}\theta_0. 
                     \nonumber
\end{IEEEeqnarray}
Finally, handover rate and mean sojourn time are obtained by
\begin{IEEEeqnarray}{rCl}
    H = \frac{1}{\mathbb{E}[S]} = \int_{v_0} \int_{\theta_0} H_{v_0,\theta_0}
      = \sqrt{\lambda} \mathbb{E}[\mathcal{F}(v,v_0,\theta_0)], 
    \label{eq:H_and_meanS}
\end{IEEEeqnarray}
where the expectation is over all the random variables (i.e., $v$, $v_0$, and $\theta_0$). 

\textbf{Special Case I}: When all drone BSs move with same velocity $v$, $\mathbb{E}[\mathcal{F}(v,v_0,\theta_0)] = \frac{4}{\pi}v$. Thus, $H=\frac{4}{\pi} \sqrt{\lambda} v$ which is equal to the handover rate of a mobile user with velocity $v$ in a single-tier network of terrestrial BSs \cite{Lin2013}.

\textbf{Special Case II}: Consider a scenario where each drone BS either moves with velocity $v>0$ or remains static\footnote{This scenario can be further extended to study the handover rate for a hybrid network of terrestrial and drone base stations.}. Let us denote the probability that a drone BS moves with $p_m$. The handover rate for this case is
$
    H = 2 \sqrt{\lambda} v p_m \left( 1 - \left( 1-\frac{2}{\pi}\right)p_m \right), 
$
where $\sqrt{\lambda} v p_m \left( 1-p_m\right)$ is the handover rate from a moving drone BS to a static drone BS which is equal to the handover rate from a static drone BS to a moving drone BS. $\frac{4}{\pi} \sqrt{\lambda} v p_m^2$ is also the handover rate from a mobile drone BS to another mobile drone BS.

Next, we solve the following optimization problem:
\begin{IEEEeqnarray}{rCl}
    & \min_{f_V}  & \quad H=\sqrt{\lambda}\mathbb{E}_{v,v_0,\theta_0}[\mathcal{F}(v,v_0,\theta_0)] \nonumber \\
    & \text{subject to} & \quad v,v_0 \sim f_V, \label{C1} \\
    && \quad \mathbb{E}[v] = c,                 \label{C2} \\
    && \quad \text{supp}(f_V) \in \mathbb{R}^+, \label{C3}
\end{IEEEeqnarray}
i.e., we want to find the speed distribution for which the handover rate is minimum. Condition \eqref{C1} indicates that $v$ and $v_0$ are two independent realizations of the distribution $f_V$. According to \eqref{C2} and \eqref{C3}, $f_V$ is a distribution with mean $c$ and positive support (it only outputs positive real numbers). In the following corollary, we  provide the solution of this optimization problem.

\begin{Corollary}
    The handover rate is minimum when all drone base stations move with speed $c$, compared to any other speed distribution with mean $c$.
\end{Corollary}
\begin{IEEEproof}
    Since $\mathcal{F}(v,v_0,\theta_0)$ is convex with respect to $v$ and $v_0$, from Jensen's inequality, we have
    \begin{IEEEeqnarray}{rCl}
        \mathbb{E}_{v,v_0,\theta_0} [\mathcal{F}(v,v_0,\theta_0)] 
        &\ge& \mathbb{E}_{v_0,\theta_0} [\mathcal{F}(\mathbb{E}[v],v_0,\theta_0)] 
        \nonumber \\
        &\ge& \mathbb{E}_{\theta_0} [\mathcal{F}(\mathbb{E}[v],\mathbb{E}[v_0],\theta_0)]. \nonumber
    \end{IEEEeqnarray}
    The equality holds only when $v$ and $v_0$ are constants. Since, $v$ and $v_0$ are two realizations of the same distribution with mean $c$, we have the equality only when $v=v_0=c$.  
\end{IEEEproof}

\section{Numerical Results}
In this section, we validate the analytical results by comparing them with simulation results. We also study the effect of mean and variance of the  speed distribution of drones on the handover rate.

\begin{figure}
	\centering
	\includegraphics[width=.25\textwidth]{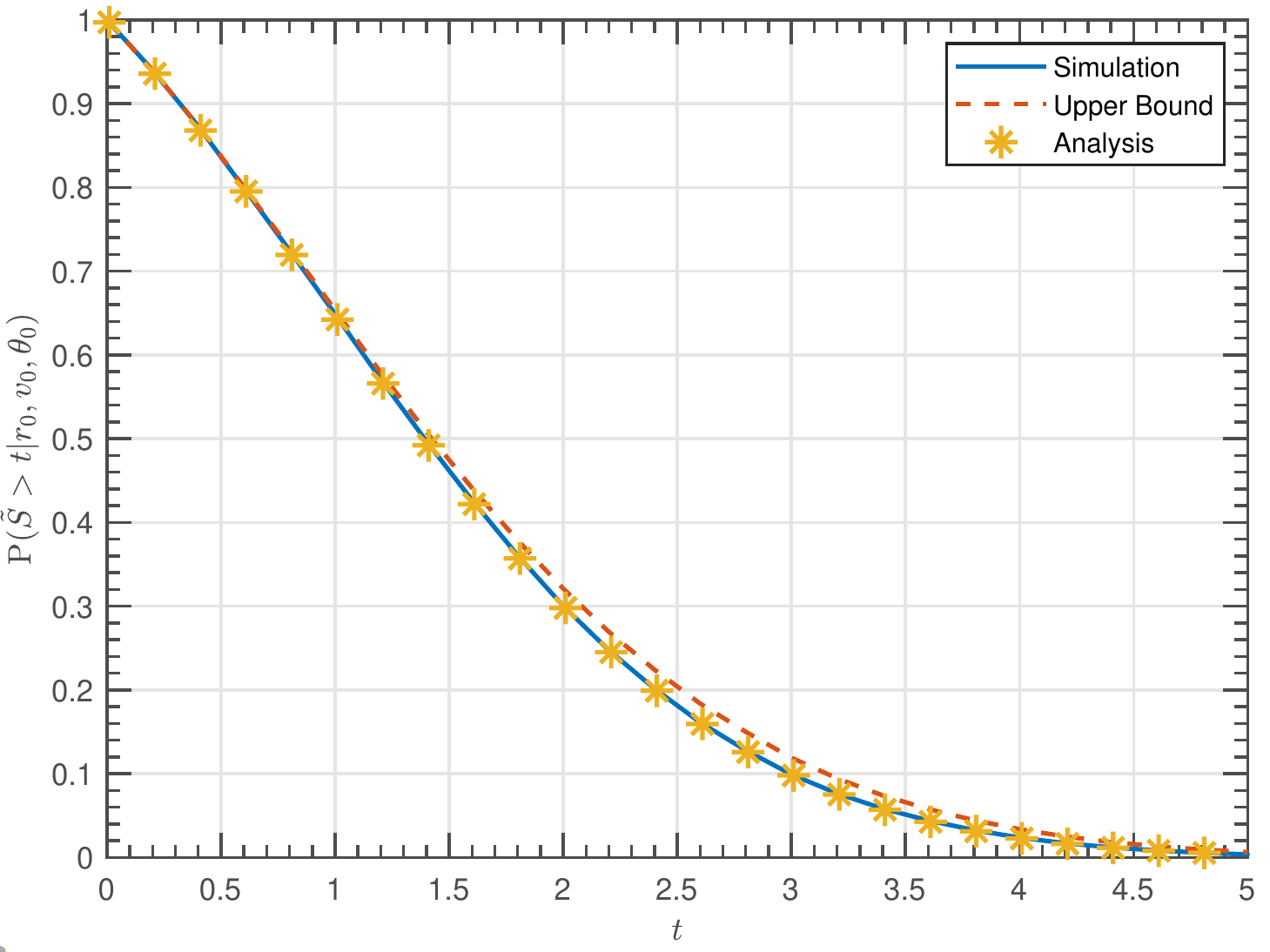}
	\caption{Conditional CCDF of time until first handover given $r_0=12$, $v_0=10$, and $\theta_0=\pi/3$. $\lambda=0.0005$. $v$ is uniformly distributed in the interval $[5,25]$. }
	\label{fig:CCDF_S_Tilde}	
\end{figure}
We compare \eqref{eq:step5} with simulation in \figref{fig:CCDF_S_Tilde}. $\tilde{S}$ denotes the time until the first handover and is greater than $s$ when there is no handover in the time interval $[0,s]$. For the scenario where different drone BSs move with the same speed, when there is a handover, the initially serving drone BS will not serve the user again (always exist a closer drone base station to the user than the old drone base station after handover). This is also proved in \textbf{Lemma 2} in \cite{Banagar2020}. Therefore, in this case, handover does not occur in the time interval $[0,s]$ if the serving base station at time $s$ is the same as the initial serving base station. However, in a scenario where drone BSs move with different speeds, a drone BS can serve a user at $t<t_1$ and $t_2<t$ (for $t_1<t_2$), while in the time interval $[t_1,t_2]$ another drone BS serves the user. In other words, $\tilde{S}$ may be less than $s$ even when the serving base stations at time $0$ and $s$ are the same. Therefore, in this case, checking for a handover event by comparing serving base stations at time $0$ and time $s$ (instead of the whole interval $[0,s]$) provides an upper bound. 
\begin{figure}
	\parbox[c]{.24\textwidth}{%
		\centerline{\subfigure[Different means and variances.]
			{\includegraphics[width=0.24\textwidth]{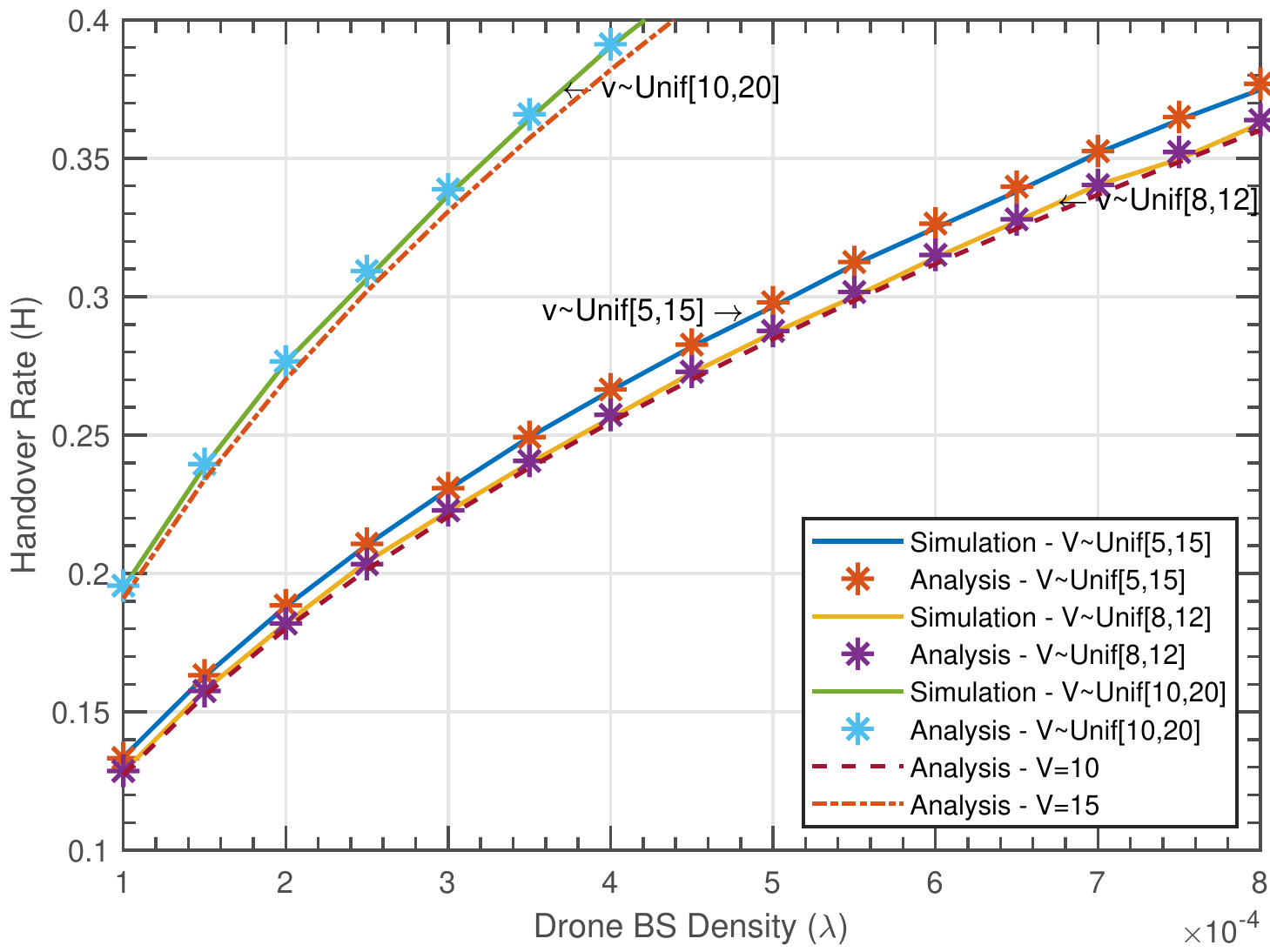}}}}
	\parbox[c]{.24\textwidth}{%
		\centerline{\subfigure[Different mobility models.]
			{\includegraphics[width=.24\textwidth]{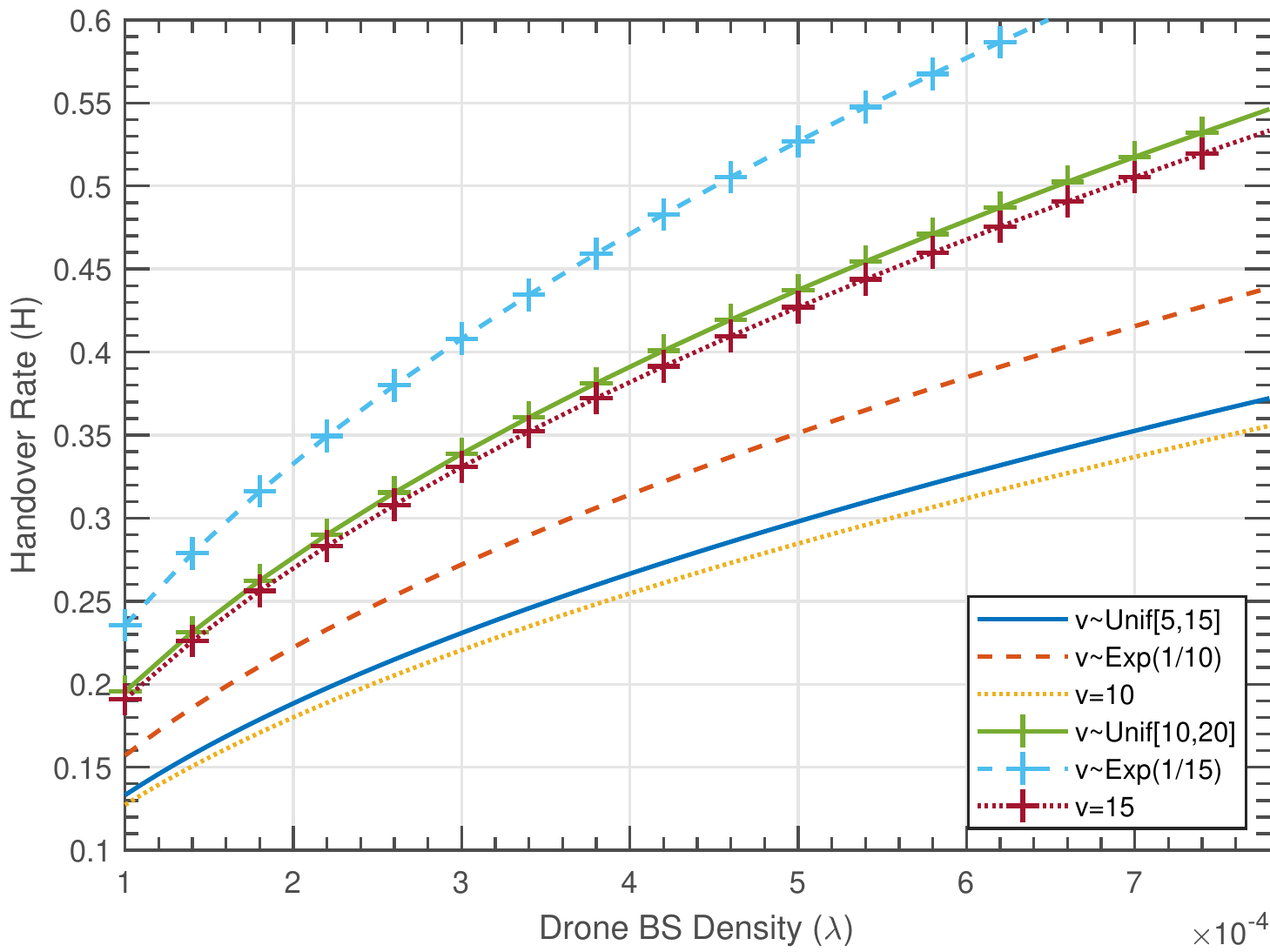}}}}
	\caption{Handover rate with respect to  drone density for different speed distributions.}
	\label{fig:HandoverRate}
\end{figure}
In \figref{fig:HandoverRate}, we show the effect of  speed distribution of drones on the handover rate for different density values of drone BSs. According to \figref{fig:HandoverRate}(a), with increasing the mean speed of drones, the handover rate increases. Also, increasing the variance of the speed distribution (while keeping its mean the same) increases the handover rate. Therefore, when all drone base stations move with same speed $v$, the handover rate is minimum, compared to the any other distribution with mean $v$. In \figref{fig:HandoverRate}(a), the handover rate  is illustrated for different mobility models. Specifically, we compare uniform distribution with exponential distribution  and deterministic distribution. These distributions are related to random walk and modified random waypoint mobility models~\cite{Tabassum}. 

\section{Conclusion}
We have derived the handover rate and mean sojourn time for a network of drone base stations. We have also shown that, handover rate is minimum when all drone base stations move with same speed (compared to any other distribution with the same mean). Although we have considered a simple network, our results can be easily extended for more complicated scenarios. Specifically,  the handover rate in a hybrid network of terrestrial and aerial base stations can be derived following the same approach.

\IEEEpeerreviewmaketitle
\bibliographystyle{IEEEtran}
\bibliography{IEEEabrv,Bibliography}

\end{document}